\documentclass[showpacs, twocolumn]{revtex4-1}
\usepackage{graphicx}
\usepackage{amsmath}
\usepackage{appendix}
\begin{document}

\title{Quantum and thermal fluctuations in two-component Bose gases}

\author{Abdel\^{a}ali Boudjem\^{a}a}
\affiliation{Department of Physics,  Faculty of Exact Sciences and Informatics, Hassiba Benbouali University of Chlef P.O. Box 78, 02000, Ouled Fares, Chlef, Algeria.}
\email {a.boudjemaa@univ-chlef.dz}

\date{\today}

\begin{abstract}

We study the effects of quantum and thermal fluctuations on Bose-Bose mixtures at finite temperature 
employing the time-dependent Hartree-Fock-Bogoliubov (TDHFB) theory.
The theory governs selfconsistently the motion of the condensates, the noncondensates and of the anomalous components on an equal footing.
The finite temperature criterion for the phase separation is established.
We numerically analyze the temperature dependence of different densities for both miscible and immiscible mixtures.
We show that the degree of the overlap between the two condensates
and the thermal clouds is lowered and the relative motion of the centers-of-mass of the condensed and thermal components is strongly damped due
to the presence of the pair anomalous fluctuations. Our results are compared with previous theoretical and experimental findings.
On the other hand, starting from our TDHFB equations, we develop a random-phase theory for the elementary excitations in a homogeneous mixture.
We find that the normal and anomalous fluctuations may lead to enhance the excitations and  the thermodynamics of the system.  

\end{abstract}

\pacs{03.75.Hh, 67.60.Bc, 03.75.Mn, 67.85.Bc} 

\maketitle

\section{Introduction} \label{Intro}

Recently, mixed ultracold quantum gases including  Bose-Bose, Fermi-Fermi, Bose-Fermi, and Bose-impurity mixtures
have attracted a great deal of interest due to their fascinating properties.
Precision measurements and novel phase transitions are among a few prominent examples provided by such mixtures.

Experimentally, binary states can be realized by using different hyperfine levels ${}^{87}$Rb \cite{Mya, Hall, Mad,Cab, Sem}, different isotopes of the same species
${}^{87}$Rb-${}^{85}$Rb \cite{Pap}, ${}^{168}$Yb-${}^{174}$Yb \cite{Sug},  
different atomic species ${}^{87}$Rb-${}^{41}$K \cite{Mog}, ${}^{87}$Rb-${}^{133}$Cs \cite{McC, Ler}, ${}^{87}$Rb-${}^{84}$Sr and ${}^{87}$Rb-${}^{88}$Sr \cite{Pasq}, 
${}^{87}$Rb-${}^{39}$K \cite{Wack}, and ${}^{87}$Rb-${}^{23}$Na \cite{Wang},  and with different statistics ${}^{6}$Li-${}^{7}$Li \cite{Igor}.
These achievements allow one to study collective modes \cite{Mad, Igor}, phase separation between the constituents \cite{Pap, McC, Wack, Wang, Tojo, Nick},  
the observation of heteronuclear Effimov resonances \cite {Bar}, and the production of polar molecules \cite{Mol}.

Theoretical investigations of degenerate binary Bose mixtures have mainly addressed the determination 
of the ground state and the density profiles of trapped systems \cite {Tin, Pu, Esry}, the stability, and the phase separation \cite{ Trip, Esry, Tim, Esry1, Rib, Jez, Svid}.
The dynamics of the center-of-mass oscillation (dipole modes) of two-component Bose-Einstein condensates (BECs) was studied analytically and numerically
by Sinatra et {\it al}. \cite{Sinatra}, whereas, the excitations of quadrupole and scissors modes have been explored by Kasamatsu et {\it al}. \cite{Kasa}. 
Furthermore, the properties of homogeneous double condensate systems were analyzed in \cite {Larsen, Bass, YNep, Sor, Tom} using the Bogoliubov theory.
%Dual condensates with dipole-dipole interactions have been also discussed within the mean field theory \cite{Wilson,Xiao, Past}. 

%On the other hand, quantum self-bound droplets have been recently predicted to occur in Bose-Bose mixtures with isotropic
%contact interactions \cite{Petrov, Petrov1} due to the competition between the negative inter-component scattering length and quantum fluctuations. 
%This novel state of matter has been most recently obseved in Barcelona group \cite {Cab} with ${}^{39}$K atoms.

At finite temperature, uniform binary Bose gases have been worked out using the Bogoliubov approach \cite{CFetter}, Hartree-Fock theory \cite{Scha} 
and a large-$N$ approximation \cite{Chien}.  The phase separation, the dynamics, and the thermalization mechanisms of trapped binary mixtures 
at finite temperatures have been also examined utilizing the local-density approximation \cite {Shi}, HFB-Popov theory \cite {Arko}, 
and the Zaremba-Nikuni-Griffin (ZNG) model \cite {Edm, Lee, Lee1}.
Very recently, effects of quantum and thermal fluctuations in a two-component Bose gas with Raman induced spin-orbit coupling 
have been analyzed using the HFB-Popov theory \cite {Hui1}.

%However, none of the above theories took into account the full dynamics of the
%anomalous average. This latter has a crucial contribution especially at intermediate temperatures \cite {}. 
%It quantifies the correlations of pairs of noncondensate atoms with pairs of condensate atoms.

%We find that the plane-wave phase is significantly broadened by thermal fluctuations

Although the above theories received great success in describing the behavior of two-component BECs, 
much remains to be investigated regarding  effects of quantum and thermal fluctuations on 
the phase separation and collective excitations of such mixtures.
The present work deals with the static and the dynamic properties of homogeneous and inhomogeneous Bose-Bose mixtures at finite temperature 
using the TDHFB theory \cite {Boudj, Boudj1, Boudj2, Boudj3, Boudj4, Boudj5, Boudj6, Boudj7, Boudj8, Boudj9, Boudj10, Boudj11}.
Our scheme provides an excellent starting point to study the dynamics of Bose systems 
and has been successfully tested against experiments in a wide variety of problems namely, collective modes, vortices, solitons and Bose polarons.  

In this paper we show that the TDHFB theory offers a rigorous and self-consistent framework to analyze the full dynamics of the two condensates, thermal clouds 
and pair anomalous correlations, including coupling between the two thermal clouds and anomalous components.
In addition, the TDHFB equations allow us to examine the role of anomalous fluctuations in the phenomenon of phase separation in trapped dual Bose condensates.
The anomalous density has a crucial contribution in the stability, excitations, superfluidity, and solitons in a single component BEC 
\cite {Boudj1, Boudj2, Griffin, Burnet, Yuk, Burnet,Giorg, Boudj10, Boudj11,Bulg}. 
Based on experimentally relevant parameters, we demonstrate that a large anomalous density may lead to a transition from miscible to immiscible regime.  
We find also that the relative motion of the centers of mass of the BECs and thermal clouds is strongly damped when the anomalous density is present  
at both zero and finite temperatures.  

In the spirit of the generalized random-phase approximation (RPA), linearized TDHFB equations are derived in order to investigate 
the collective excitations in a homogeneous mixture at finite temperature. The developed theory can be referred to as the TDHFB-RPA. 
Neglecting the intraspecies interactions and keeping only terms of second order in coupling constants, the TDHFB-RPA reduces to the 
finite temperature second-order Beliaev theory \cite{Griffin}.
The ultraviolet divergence of the anomalous averages is properly regularized obtaining useful analytical expression.
Effects of quantum and thermal fluctuation corrections in the excitations and the thermodynamics are deeply analyzed.

The rest of the paper is organized as follows. In Sec.\ref {Model}, we outline the general features of the TDHFB equations derived for 
binary Bose condensates. 
We discuss also the main hindrances encountered in our model and present the resolution of these problems.
The finite temperature stability condition of the mixture is accurately identified. 
Section \ref{TBBM}  deals with harmonically trapped Bose-Bose mixtures and is divided into two subsections related to several subjects.
Section \ref{DP}  is devoted to solving our equations numerically in a three-dimensional (3D) case and 
analyzing the profiles of the condensed, noncondensed, and anomalous densities in terms of temperatures  for miscible and immiscible mixtures.
We will look at in particular how the anomalous fluctuations enhance the degree of the overlap between both  the condensates and thermal clouds.
It is found that the phase separation between the condensates is suppressed  as the temperature is increased in good agreement with the HFB-Popov approximation \cite{Arko}.
In Sec.\ref{TE}  we analyze the dynamics of two trapped  BECs in the presence of
the thermal cloud and the pair anomalous correlation at both zero and finite temperatures.
We relate our findings to those of previous experimental and theoretical treatments.
In Sec.\ref{HM}  we solve our TDHFB equations  to second order in the interaction coupling constants for uniform mixture at finite temperature using the generalized RPA.
We show that the TDHFB-RPA method constitutes a finite-temperature extension of the Beliaev approximation discussed in a single component Bose condensed gas 
with contact interaction \cite{Griffin, Beleav} and dipole-dipole interactions \cite{Boudj2015}.
Meaningful analytical expressions are obtained for the excitations spectrum, the condensed depletion, the anomalous density, the equation of state (EoS) and the ground state energy.
Finally, we conclude in Sec.\ref{concl}.

\section {TDHFB Theory} \label{Model}

We consider weakly interacting two-component BEC  with the atomic mass $m_j$ confined in external traps $V_j ({\bf r})$.
The many-body Hamiltonian describing such mixtures reads 
\begin{align} \label{eq4}
\hat H &= \sum_{j=1}^2\int d{\bf r} \, \hat\psi_j^\dagger ({\bf r}) \left[h_j^{sp} +\frac{g_j}{2} \hat \psi_j^\dagger ({\bf r})\hat \psi_j ({\bf r})\right]\hat\psi_j ({\bf r}) \\ \nonumber
&+g_{12}\int d{\bf r} \, \hat\psi_2^\dagger ({\bf r}) \hat \psi_2({\bf r})\hat \psi_1^\dagger ({\bf r}) \hat\psi_1({\bf r}),
\end{align}
where  $\hat\psi_j^\dagger$ and  $\hat\psi_j$ are the boson destruction and creation field operators, respectively, satisfying the usual canonical commutation rules 
$[\hat\psi_j({\bf r}), \hat\psi_j^\dagger (\bf r')]=\delta ({\bf r}-{\bf r'})$.
The single particle Hamiltonian is defined by $h_j^{sp}=-(\displaystyle\hbar^2/\displaystyle 2m_j) \Delta + V_j$.
The coefficients $g_j=(4\pi \hbar^2/m_j) a_j$ and $g_{12}=g_{21}= 2\pi \hbar^2 (m_1^{-1}+m_2^{-1}) a_{12}$ with 
$a_j$ and $a_{12}$ being the intraspecies and the interspecies scattering lengths, respectively. 

At finite temperature, we usually perform our analysis in the mean-field framework relying on the TDHFB equations.
For Bose mixtures, the TDHFB equations are given by \cite {Boudj5, Boudj6}
\begin{equation} \label {TDH1}
i\hbar \frac{d  \Phi_j}{d t} =\frac{d{\cal E}}{d \Phi_j},
\end{equation}
\begin{equation} \label {TDH2}
i\hbar \frac{d \rho_j}{d t} =-2\left[\rho_j, \frac{d{\cal E}}{d\rho_j} \right],
\end{equation}
where ${\cal E}=\langle \hat H\rangle$ is the energy of the system.
In Eqs.(\ref{TDH2}), $\rho_j ({\bf r},t)$ is the single particle density matrix of a thermal component  defined as
$$
\rho_j=\begin{pmatrix} 
\langle \hat{\bar{\psi}}^\dagger\hat{\bar{\psi}}\rangle & -\langle\hat{\bar{\psi}}\hat{\bar{\psi}}\rangle\\
\langle\hat{\bar{\psi}}^\dagger\hat{\bar{\psi}}^\dagger\rangle& -\langle\hat{\bar{\psi}}\hat{\bar{\psi}}^\dagger\rangle
\end{pmatrix}_j,
$$
where $\hat{\bar \psi}_j({\bf r})=\hat\psi_j({\bf r})- \Phi_j({\bf r})$ is the noncondensed part of the field operator with $\Phi_j({\bf r})=\langle\hat\psi_j({\bf r})\rangle$ 
being the condensate wave-function. 
Equations (\ref{TDH1}) and (\ref{TDH2}) are  obtained using the Balian-V\'en\'eroni variational principle \cite{BV} 
that optimizes a generating functional related to the observables of interest.
The single component BEC version of Eqs.(\ref{TDH1}) and (\ref{TDH2}) was derived in \cite{Ben}.
%This latter is based on the minimization of an action  which involves two variational objects : one related to the observables of interest and the other is akin to a density matrix.
%In our case, we choose a Gaussian time-dependent density operator ${\cal D}(t)$.
%For any operator one has  $\langle \hat O \rangle ({\bf r}, t)=\hbox{Tr}\,\hat O({\bf r})\, {\cal D}(t) /{\cal Z}(t)$,
%where ${\cal Z}(t)=\hbox{Tr}\,{\cal D}(t)$, is the partition function \cite{BV, Ben}.

An important feature of the TDHFB  formalism is that it allows unitary evolution of  $\rho_j$.
Then it follows that
\begin{align} \label{Invar}
\rho_j (\rho_j +1)= ( I_j-1)/4,
\end{align} 
where $I$ is often known as the Heisenberg invariant \cite{Cic,Ben, Boudj}. It represents the variance of the number of noncondensed particles. 
For pure state and at zero temperature, $I=1$.

%The BV variational principle enables us to express the von Neumann entropy as a function of the density operator ${\cal D}$ as $S = −\hbox{Tr}\,{\cal D} \ln{\cal D}$. 
%Recall that $S$ is a measure of the distribution of the eigenvalues of ${\cal D}$. 
%It may play also a crucial role in quantifying the entanglement between the two BECs. 
%It can be written also in terms of the invariant $I$ as \cite{Cic}
%\begin{equation} \label{Entpy}
%S= \int \frac{d {\bf k}} {(2\pi)^3}  \bigg[\frac{1}{2} \sqrt{I} \ln \left( \frac{\sqrt{I}+1} {\sqrt{I}-1} \right) + \frac{1}{2} \ln (I-1) \bigg].
%\end{equation}
%When the states of the two species have equal probability to be occupied, $S$ reaches its maximal value.

The total energy can be easily computed yielding: 
%\begin{widetext}
\begin{align}  \label{egy}
{\cal E} &= \sum_{j=1}^2 \bigg[ \int d{\bf r} \, \left( \Phi_j^*  h_j^{sp} \Phi_j + \hat{\bar \psi}_j^\dagger  h_j^{sp}  \hat{\bar \psi}_j  \right)  \\
&+ \frac{g_j}{2} \int d{\bf r} \bigg( n_{cj}^2+ 4\tilde n_j n_{cj} +2\tilde n_j^2 +|\tilde m_j|^2 \nonumber\\
&+ \tilde m_j^*\Phi_j^2 + \tilde m_j {\Phi_j^*}^2 \bigg) \bigg] \nonumber \\
&+ g_{12}\int d{\bf r}\, (n_{c1}+\tilde n_1)  (n_{c2}+\tilde{n}_2), \nonumber
\end{align}
%\end{widetext}
where  $n_{cj}=|\Phi_j|^2$ is the condensed density, $\tilde n_j=\langle\hat{\bar {\psi}}_j ^\dagger\hat{\bar {\psi}}_j\rangle$ is the noncondensed density, 
and $\tilde m_j= \langle\hat {\bar {\psi}}_j\hat{\bar {\psi}}_j\rangle$ is the anomalous density.

Upon introducing the expression (\ref{egy}) into Eqs.(\ref {TDH1}) and (\ref {TDH2}), one obtains the explicit TDHFB equations for the two-component BECs
\begin{subequations}\label {T:DH}
\begin{align} 
i\hbar \dot{\Phi}_j & = \left[ h_j^{sp}+g_j (n_{cj}+2\tilde n_j) + g_{12}  n_{3-j} \right]\Phi_j  \label{T:DH1}  \\
&+ g_j\tilde m_j \Phi_j^{*} , \nonumber  \\  
i\hbar \dot{\tilde n}_j &= g_j\left(\tilde m_j^{*}\Phi_j^2-\tilde m_j {\Phi_j^{*}}^2\right) ,  \label{T:DH2} \\ 
i\hbar \dot{\tilde m}_j &=  4\left[ h_j^{sp}+2g_j n_j+\frac{ g_j }{4} \left (2\tilde n_j +1\right)+g_{12} n_{3-j} \right] \tilde m_j  \nonumber \\
&+g_j (2\tilde n_j +1)\Phi_j^2  \label{T:DH3},
\end{align}
\end{subequations}
where $n_j=n_{cj}+\tilde n_j $ is the total density.
Setting $g_{12} =0$, one recovers the usual TDHFB equations \cite {Boudj, Boudj1, Boudj2, Boudj3, Boudj4, Boudj7, Boudj10, Boudj11} 
describing a degenerate Bose gas at finite temperature.
In a highly imbalanced mixture where $g_1=0$ or $g_2=0$, Eqs.(\ref{T:DH}) coincide with our TDHFB equations 
recently employed in Bose-polaron systems \cite {Boudj5, Boudj6, Boudj8, Boudj9}. 
For $\tilde n_j=\tilde m_j=0$,  they reduce to the coupled Gross-Pitaevskii (GP) equations for binary condensates at zero temperature.
In the case of a Fermi-Fermi mixture, Eq.(\ref{T:DH1}) has no analog, while Eqs.(\ref{T:DH2}) and (\ref{T:DH3}) stand for the Hartree-Fock and the gap equations, respectively.
In the semiclassical limit, the TDHFB is equivalent to the collisionless Boltzmann equation for the particle distribution function \cite{Giorg}.

Indeed, the TDHFB theory, as the standard HFB approximation, runs into trouble. 
The first problem is the destruction of the gaplessness of the TDHFB theory due to the inclusion of  the anomalous density 
signaling that the theory satisfies neither the Hugenholtz-Pines theorem \cite{HP} nor the Nepomnyashchy identity \cite {NP}. 
Secondly,  the anomalous pair average which in general leads to a double counting of the interaction effects is ultraviolet divergent \cite{DStoof}. 
Physically this comes from the contact interaction potential, which treats collisions of different momenta with the same probability.
To reinstate the gaplessness of the spectrum, one should renormalize the intraspecies coupling constants $g_j$ 
following the procedure outlined in Refs \cite{Burnet, Boudj, Boudj4, Boudj6} for a single BEC. This gives
\begin{align} \label{Ren}
\bar g_j= g_j(1+\tilde m_j/\Phi_j^2).
\end{align} 
Despite the dilute nature of the system,  the spatially dependent effective interaction $\bar g_j $ may modify the static and the dynamics of the mixture.
Furthermore, $\bar g_j$ have substantial implications for the stability condition. 
It is worth noticing that this technique renders the TDHFB equations (\ref{T:DH}) gapless but leaves the anomalous density divergent as we shall see in Sec.\ref{HM}.\\
Given Eq.(\ref{Ren}), the renormalized TDHFB equations read
\begin{subequations}\label {RT:DH}
\begin{align} 
i\hbar \dot{\Phi}_j & = \left[ h_j^{sp}+ \bar g_j n_{cj}+2g_j \tilde n_j + g_{12}  n_{3-j} \right]\Phi_j,  \label {RT:DH1}  \\
i\hbar \dot{\tilde m}_j &=  4\left[ h_j^{sp}+2g_j n_j+G_j \left (2\tilde n_j +1\right)+g_{12} n_{3-j} \right] \tilde m_j   \label {RT:DH2},
\end{align}
\end{subequations}
where $G_j$ is related to $\bar g_j$ via $G_j=g_j \bar g_j /4(\bar g_j-g _j)$.
Equations (\ref{RT:DH}) are appealing since they permit us to study the behavior of the thermal cloud and the pair anomalous density of Bose-Bose atomic mixtures at any temperature.
It is easy to check that they satisfy the energy and number conserving laws.

Equilibrium states  can be readily determined via the transformations:
$\Phi_j ({\bf r},t)= \Phi_j ({\bf r}) \exp(-i \mu_j t/\hbar)$ and $\tilde m_j ({\bf r},t)= \tilde m_j ({\bf r}) \exp (-i \mu_j t/\hbar)$, 
where $\mu_j$ are chemical potentials related with each components. Here $\mu_j$ must be calculated self-consistently employing the normalization condition 
$N_j=\int n_j d {\bf r}$, where $N_j=N_{cj}+\tilde N_j$ is the single condensate total number of particles with 
$N_{cj}=\int n_{cj} d {\bf r}$ and $\tilde N_j=\int \tilde n_j d {\bf r}$ being respectively, the condensed and noncondensed number of particles in each component.

A useful relation between the normal and anomalous densities can be given via Eq.(\ref {Invar}) 
\begin{equation}  \label{Inv1}
I_j= (2\tilde n_j+1)^2- 4|\tilde m_j |^2.
\end{equation}
This equation clearly shows that when $I \rightarrow $1 or equivalently  $T\rightarrow 0$, the absolute value of the anomalous density is larger than the noncondensed density.
%By differentiating Eq.(\ref{Inv1}), one can see that neglecting $\Delta \tilde m$ does not mean omitting $ \Delta \tilde n$,  justifying the use of the 
%Thomas-Fermi for the anomalous component \cite {Boudj4}.
In the quasiparticle space, one has  $\tilde n_j=\sum_k \left[v_{kj}^2+(u_{kj}^2+v_{kj}^2)N_{kj}\right]$ and $\tilde m_j=-\sum_k \left[u_{kj} v_{kj} (2N_{kj}+1)\right]$,
where $N_{kj}=[\exp(\varepsilon_{kj}/T)-1]^{-1}$ are occupation numbers for the excitations and 
$ u_{kj},v_{kj}=(\sqrt{\varepsilon_{kj}/E_k}\pm\sqrt{E_{kj}/\varepsilon_{kj}})/2$ are the Bogoliubov functions with $E_{kj}$ being the energy of the free particle 
and $\varepsilon_k$ is the excitations energy. 
Combining the expressions of $\tilde m_j$ and $\tilde n_j$ and using the fact that $2N (x)+1= \coth (x/2)$, we obtain
$I_{kj} =\text {coth} ^2\left(\varepsilon_{kj}/2T \right)$.
For a noninteracting Bose gas where the anomalous density vanishes, $I_{kj} =\text {coth} ^2\left(E_{kj}/2T\right)$ \cite{Boudj10}.
For an ideal trapped case, the Heisenberg invariant keeps the same form as Eq.(\ref{Inv1}) 
with only setting $ \varepsilon_{kj} \rightarrow \varepsilon_j ({\bf p, r})=p_j^2/2m +V_j({\bf r})$,
which can be calculated within the semiclassical approximation. 
Equation (\ref{Inv1}) allows us to determine in a very convenient manner the critical temperatures of the mixture.

%{\bf Near the transition temperature where the system surive only with noncondensed part i.e. ($n_c=\tilde m=0$), 
%Eq. (\ref{RT:DH2}) corresponds to the linearized time-dependent Hartree-Fock (TDHF) equation.}

\section {Trapped Bose-Bose mixture} \label{TBBM}

\subsection {Density profiles} \label{DP}

As a starting point,  it is useful to establish the stability condition.
Working in the Thomas-Fermi (TF) approximation which consists in neglecting the kinetic terms in Eqs.(\ref {RT:DH1}) and valid for large number of particles. 
%Of course, the TF is not appropriate for thermally excited atoms due to quick variation of their kinetic energy.
The resulting equations for the condensed density distributions $n_{c1}$ and $n_{c2}$ are given by
\begin{align}
n_{c1}&= \frac{\Delta}{\bar g_1 (\Delta -1)} \bigg [ \mu_1-V_1-2g_1\tilde n_1-g_{12} \tilde n_2  \label{TF1} \\
&- \frac{g_{12}}{\bar g_2} \left(\mu_2-V_2-2g_2 \tilde n_2-g_{12} \tilde n_1 \right) \bigg], \nonumber \\
n_{c2}&= \frac{\Delta}{\bar g_2 (\Delta -1)}  \bigg [\mu_2-V_2-2g_2\tilde n_2-g_{12} \tilde n_1  \label{TF2} \\
&-\frac{g_{12}}{\bar g_1} \left(\mu_1-V_1-2g_1 \tilde n_1-g_{12} \tilde n_2 \right) \bigg], \nonumber 
\end{align}
where $\Delta=\bar g_1 \bar g_2/ g_{12}^2$ is often known as the miscibility parameter.
In our case the mixture can be miscible if $\Delta>1$ or immiscible when $\Delta<1$. 
The transition between the two regimes was previously observed in Bose-Bose mixtures in different spin states \cite{Tojo, Nick}, 
Bose-Bose mixtures of two Rb isotopes  \cite{Pap}, and in heteronuclear Bose-Fermi mixtures \cite{Osp, Zac}.
If $g_{12}=0$ and one component vanishes (say $n_1=0$) in a certain space region,  Eqs.(\ref{TF1})  and (\ref{TF2}) simplify to the one-component TF equation, namely
$n_{c2}= (\mu_2-V_2-2 g_2\tilde n_2)/\bar g_2$. For $\tilde n_j=\tilde m_j=0$, they reduce to the usual TF equations at zero temperature.
Inspection of Eqs.(\ref{TF1})  and (\ref{TF2}) suggests that the stability of the mixture merely requires the conditions:
\begin{equation}  \label{Stab}
\bar g_1 >0, \;\;\;\;\;\, \bar g_2 >0,  \;\;\;\ \text{and} \;\;\;\;\;\,  \Delta>1.
\end{equation}
In the limit $\tilde m_j/n_{cj} \ll 1$, the conditions (\ref{Stab}) reduce to the standard stability conditions at zero temperature namely $g_1 g_2>g_{12}^2$.
For $\tilde m_j/n_{cj} >1$, the system becomes strongly correlated. This means that at finite temperature, the stability criterion of the mixture requires 
the inequality $-1<\tilde m_j/n_{cj} < 1$.
If $\Delta<1$ and $g_{12} <0$, the gas is unstable whereas, for $g_{12} >0$, the two components do not overlap with each other (separated solutions).
One of the most important feature arising from our formula (\ref{Stab}) is that when $\tilde m_j$ is large, the mixture undergoes a transition from 
miscible to immiscible phase.   

In order to illustrate our approach,  we consider the ${}^{133}$Cs-${}^{87}$Rb mixture confined in a spherical trap $V_j (r)= m_j \omega_j^2 r^2/2$ 
with trap frequency $\omega_{\text{Cs}}=\omega_{\text{Rb}}= 2\pi \times 270$Hz. 
Notice that our theory can adequately treat all the existing mixtures.
The intraspecies scattering lengths are: $a_{\text{Cs}}= 280\, a_0$ and $a_{\text{Rb}}=104\, a_0$ with $a_0$ being the Bohr radius, 
the interspecies scattering lengths can be adjusted by means of a Feshbach resonance,
and particle numbers $N_{\text{Cs}} = N_{\text{Rb}} = 5 \times 10^4$. The  critical temperature for an ideal gas is about $T_c^0= 450$ nK.

\begin{figure}
\centerline{
\includegraphics[scale=0.5]{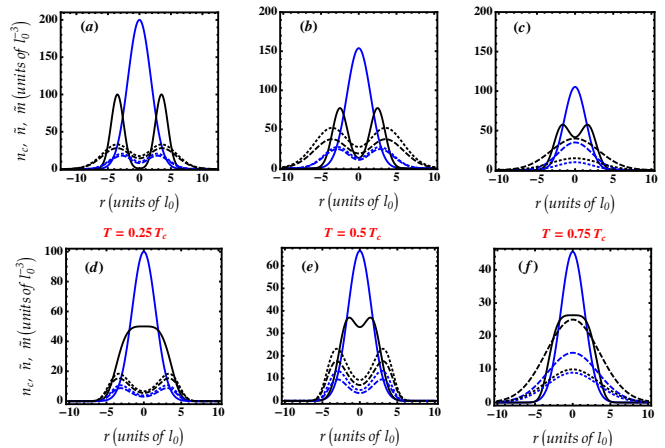}}
 \caption{ (Color online)  Top row:  density profiles of an immiscible mixture of ${}^{87}$Rb (blue) and ${}^{133}$Cs (black) atoms at different temperatures for $\Delta=0.9$.
Solid lines: condensed density. Dashed lines: thermal cloud density. Dotted lines: anomalous density. 
The thermal cloud and the anomalous densities have been amplified by 10 times for clarity.
Bottom row: same as top panels but for a miscible mixture  for $\Delta=2.5$.}
\label{DF} 
\end{figure}

Figure \ref{DF}  clearly shows that at low temperature,  the Rb has a higher peak density and narrower width while the Cs atoms are pushed towards the outer
part forming a shell structure around the Rb BEC [see Fig.\ref{DF} (a)].
Such a symmetrical demixed phase, can be understood from the fact that the Rb sustains an extra confinement from the Cs shell surrounding it,
i.e. originating from the coupling term $g_{12} n_{3-j}$ in Eq.(\ref{RT:DH}).  
For phase separation, the TF approximation becomes less satisfactory.
In this case, the Rb is located at the phase boundary, such that the density distribution varies fast in space and the
kinetic terms cannot be omitted \cite {Pu}.

At $T=0.5\, T_c$, the two components start to overlap with each other
and the overlap region is broadened with temperature [see Fig.\ref{DF}(b)].
At $T \geq 0.75\,T_c$ where the binary condensates survive with significant thermal clouds,
the mixture becomes completely immiscible as is depicted in Fig.\ref{DF} (c).
This suppression of the phase separation which has been predicted also by the HFB-Popov theory \cite{Arko}, 
can be attributed to the strong effects of thermal fluctuations.
We observe from the same figure that the anomalous density is larger than the noncondensed density at low temperature, it reaches its maximum at intermediate temperatures,
and vanishes near the transition similar to the case of a single component. 
Indeed, this behavior remains valid irrespective of the mixture whether miscible or immiscible. 
At higher temperature, both $\tilde n$ and $\tilde m$ have a Gaussian shape since the system becomes ultra-dilute \cite {Boudj2}.

Figure \ref{DF} (d) shows that at $T=0.25\,T_c$, both species overlap at the trap center. 
Remarkably, with an increase in temperature ($T=0.5\, T_c$), the mixture becomes partially immiscible [see Fig.\ref{DF} (e)].
As we have foreseen above, this phase transition is most likely due to the inclusion of the anomalous correlation 
which has a significant effect at this range of temperature.
At $T \geq 0.75\, T_c$, the mixture restores its miscibility due to the weakness of  $\tilde m$ [see Fig.\ref{DF} (f)].

\begin{figure}
\centerline{
\includegraphics[scale=0.5]{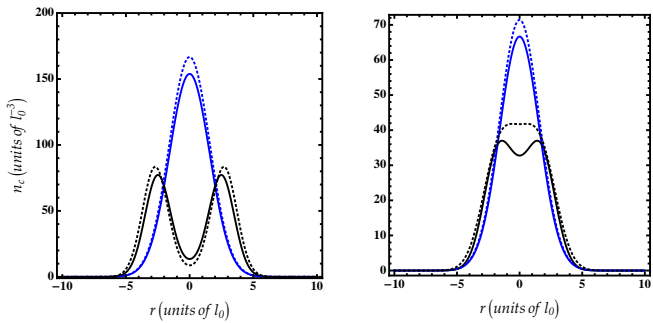}}
 \caption{ (Color online)  Condensed density for $\Delta=0.9$ (left) and $\Delta=2.5$ (right) at $T=0.5\,T_c$.
Solid lines: our predictions. Dotted lines: the results of the HFB-Popov theory. Parameters are the same as in Fig.\ref{DF}.}
\label{DF1} 
\end{figure}

In Fig.\ref{DF1} we compare our results for the condensed density with the HFB-Popov calculations. As is clearly seen, the presence of the anomalous density 
leads to reduction of the condensed density and the degree of the overlap between the two condensates. 
This is owing to the mutual interaction between condensed atoms on the one hand and the condensed atoms and noncondensed atoms on the other.

\subsection {Dynamics of spatial separation} \label{TE}

\begin{figure}
\centerline{
\includegraphics[scale=0.8]{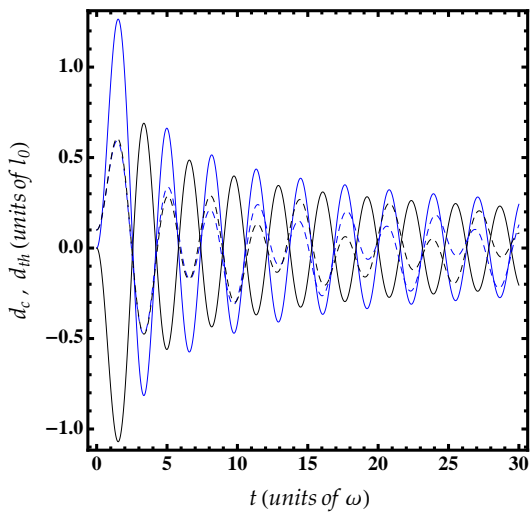}}
 \caption{ (Color online)  Mean separation between the condensates (solid line) and thermal clouds (dotted lines) versus time in isotropic traps 
for $z_0/l_0=0.3$ and  $\Delta=2.5$ at $T=0.5 T_c$.}
\label{MSD} 
\end{figure}

\begin{figure}
\centerline{
\includegraphics[scale=0.8]{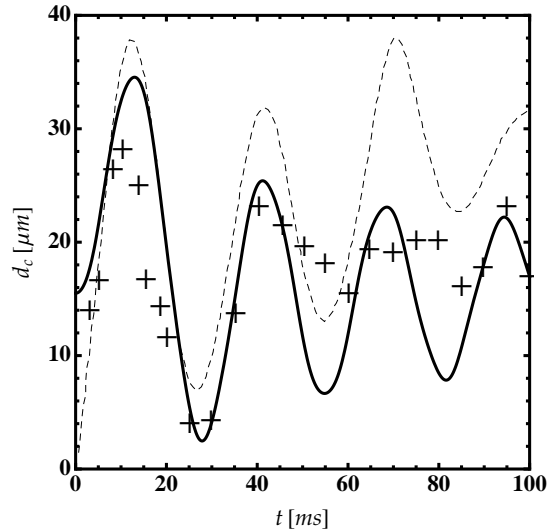}}
 \caption{ Mean separation between the condensates versus time in isotropic traps for ${}^{87}$Rb atoms, with 
the scattering lengths are  $ a_1: a_{12} :a_2 ::1.03:1:0.97$, with the average of the three being 55 \AA \,\cite{Hall}. 
Solid line : our predictions. Dashed line: theoretical results of \cite{Sinatra}. Plus: JILA experiment \cite{Hall}. 
These data were taken after 22 ms after switching off the trapping potential.}
\label{MSD1} 
\end{figure}

Let us consider the evolution of dual condensates in the presence of their own thermal clouds and anomalous components 
confined in a spherically symmetric trap which at $t=0$ its centers for the first and second component are, respectively, displaced 
along the $z$-axis by distances $\pm z_0/2$.
The separation is assumed to be small compared to the TF radii.
The time dependence of the mean separation between the two condensates is given  by
\begin{equation}  \label{MdS}
d_c (t) = \int d {\bf r} z  \left [n_{c1} ({\bf r}, t) -n_{c2} ({\bf r},t) \right],
\end{equation}
while the mean separation between the two thermal clouds reads
\begin{equation}  \label{Mdth}
d_{th} (t) = \int d {\bf r} z  \left [ \tilde n_1 ({\bf r}, t) - \tilde n_2 ({\bf r},t) \right].
\end{equation}

The numerical integration of our Eqs.(\ref{MdS}) and (\ref{Mdth}) shows that  at finite temperature, 
the relative motion of the centers-of-mass of the BECs and thermal clouds in a miscible mixture is strongly damped in particular at long time scales (see Fig.\ref{MSD}).
Such a damping, which has also been predicted by the ZNG theory \cite{Lee,Lee1},  is caused by condensate-condensate, 
condensate-thermal, and thermal-thermal interactions. 
The intra- and inter-component anomalous pair correlations may play also a crucial role for the appearance of the aforementioned 
damping of oscillations especially at $T \approx 0.5\,T_c$.
At fixed temperature, the damping of the oscillations of the mean separation between the condensates, and the thermal clouds becomes more and more strong
for a large displacement, $z_0$, regime.
One can expect that the same behavior persists in the immiscible mixture but with larger oscillation amplitudes.

To better understand the impact of the pair anomalous density on the damping mechanism, we compare 
our predictions for the relative motion of the centers-of-mass of the two BECs with the experimental measurements of \cite{Hall} 
and the theoretical results of \cite{Sinatra} based on the GP equation.
As is clearly visible from Fig.\ref{MSD1}, the curves of our TDHFB model improve the theoretical result of Ref. \cite{Sinatra} especially at large time scale. 
This correction makes our theory in good agreement with the JILA experiment \cite{Hall}.
The difference between the two models may be justified in terms of the significant contribution of the anomalous density which causes a huge loss
of atoms during the oscillation even at zero temperature.

\section {Homogeneous Bose-Bose mixture } \label{HM}

%\subsection{Excitations}

%In this section we will show that our formalism can reproduce the mean-field result based on the usual HFB approximation.

In this section we analyze the elementary excitations in a homogeneous mixed Bose gas, where $V_j(r)=0$ using the generalized RPA \cite{Boudj1, Giorg, Xia}.
This latter consists of imposing small fluctuations of the condensates, the noncondensates, and the anomalous components, respectively, as: 
%The essential idea of the RPA is that the gas responds as a reference gas to self-consistent dynamical potentials [8].
$\Phi_j = \sqrt{n_{cj}}+\delta \Phi_j $, $\tilde n_j=\tilde n_j+\delta \tilde n_j$,  and $\tilde m_j=\tilde m_j+\delta \tilde m_j$,  
where $\delta \Phi_j \ll \sqrt{n_{cj}}$, $\delta \tilde n_j \ll \tilde n_j$, and $\delta \tilde m_j \ll \tilde m_j$. 
Thus, we obtain the TDHFB-RPA equations 
\begin{align} 
i\hbar \delta  \dot \Phi_j & = \left[ h_j^{sp}+ 2\bar g_j n_{cj}+2g_j \tilde n_j + g_{12} n_{3-j} \right] \delta \Phi_j  \label {RPA1} \\ 
    &+\bar g_j n_{cj} \delta \Phi_j^*+ 2g_j \sqrt{n_{cj}} \delta \tilde n_j  + g_{12}\sqrt{n_{c{3-j}}} \delta \tilde n_{3-j} \nonumber \\
&+ g_{12} \sqrt{ n_{cj} n_{c{3-j}} } (\delta \Phi_{3-j}+\delta \Phi_{3-j}^*),  \nonumber 
\end{align}
and 
\begin{align} 
i\hbar \delta  \dot{\tilde m}_j &=  4\left[ h_j^{sp}+2g_j n_j+G_j (2\tilde n_j +1)+g_{12} n_{3-j} \right] \delta\tilde m_j   \label {RPA2} \\ 
&+ 8g_j \tilde m_j \left[ \sqrt{ n_{cj}}  (\delta \Phi_j+ \delta \Phi_j^*)+ \delta \tilde n_j+ (G_j/g_j) \delta \tilde n_j \right] \nonumber\\
&+ g_{12}\tilde m_j \left[\sqrt{ n_{c{3-j}}}  (\delta \Phi_{3-j}+\delta \Phi_{3-j}^*) +\delta \tilde n_{3-j} \right]. \nonumber
\end{align}
Here we recall that $\delta \tilde n_j$ and $\delta \tilde m_j$ are related with each other through (\ref{Inv1}).
Remarkably, Eqs.(\ref{RPA1})  and (\ref{RPA2}) contain a class of terms beyond second order.
They can be regarded as a natural extension of the HFB-RPA \cite{Xia} theory developed for the single component BEC. 
If one neglects the anomalous density,  the TDHFB-RPA equations reduce to the HFB-Popov-RPA equations.

Since we restrict ourselves to second order in the coupling constants, one must retain in Eqs.(\ref{RPA1})  and (\ref{RPA2}) 
only the terms which describe the coupling to the condensate and neglect all terms associated with 
fluctuations  $\delta \tilde n$ and $\delta \tilde m$ \cite {Giorg}.
In fact, this assumption  is relevant to ensure the gaplessness of the spectrum.
Writting the field fluctuations associated to the condensate in the form 
$\delta \Phi_j ({\bf r},t)= u_{jk}  e^{i {\bf k \cdot r}-i\varepsilon_k t/\hbar}+v_{jk} e^{i {\bf k \cdot r}+i\varepsilon_k t/\hbar}$,  
we obtain the second order coupled TDHFB-de Gennes equations for the quasiparticle amplitudes $u_{kj}$ and $v_{kj}$ :
\begin{equation} \label{BdG}
\begin{pmatrix} 
{\cal L}_1 & {\cal M}_1 &  {\cal A}  &  {\cal A}
\\
  {\cal M}_1 & {\cal  L}_1 &  {\cal A} &  {\cal A}
\\
  {\cal A} &  {\cal A}& {\cal L}_2 &  {\cal M}_2
\\
 {\cal A} &  {\cal A} &  {\cal M}_2 & {\cal  L}_2
\end{pmatrix}\begin{pmatrix} 
u_{1k} \\ v_{1k} \\ u_{2k}  \\ v_{2k} 
\end{pmatrix}=\varepsilon_k \begin{pmatrix} 
u_{1k} \\ -v_{1k}  \\ u_{2k}  \\ -v_{2k}
\end{pmatrix},
\end{equation} 
where $\int d {\bf r} [u_j^2( {\bf r})- v_j^2({\bf r})]=1$, 
${\cal L}_j = E_k+ 2 \bar g_j n_{cj}+ 2 g_j \tilde n_j + g_{12} n_{3-j} -\mu_j$,  ${\cal M}_j= \bar g_j n_{cj}$, 
${\cal A}=g_{12}\sqrt{n_{c1} n_{c2} }$, $\varepsilon_k$ is the Bogoliubov excitation energy and 
$E_k=\hbar^2k^2/2m$ is the kinetic energy which is the same for both species since we consider equal masses ($m_1=m_2=m$).  
For $g_{12}=0$, Eqs.(\ref{BdG}) coincide with the finite temperature second-order equations obtained by Shi and Griffin using diagrammatic methods \cite{Griffin}
and with the finite temperature time-dependent mean-field scheme proposed by Giorgini \cite{Giorg}.
At zero temperature they correspond to the well-known second-order Beliaev's results \cite{Beleav} discussed in a single component Bose condensed gas over six decades ago,
while at high temperature our second-order coupled TDHFB-de Gennes equations reproduce those derived by Fedichev and Shlyapnikov \cite{FedG} 
employing Green's function perturbation scheme.\\
The chemical potentials turn out to be given as
\begin{equation} \label{EoS}
\mu_j= \bar g_j n_{cj} + 2 g_j \tilde n_j + g_{12} n_{3-j},
\end{equation}
Inserting Eq.(\ref{EoS}) into (\ref{BdG}), one obtains the following Bogoliubov spectrum composed of two branches:
\begin{equation} \label {Bog}
\varepsilon_{k+}= \sqrt{E_k^2+2E_k \mu_+}\,,  \,\,\,\,\,\,\,\,\,\, \varepsilon_{k-}= \sqrt{E_k^2+2E_k \mu_-}\,, 
\end{equation}
where
%$$\mu_{+,-}=  \frac{1}{2} \bigg[ \bar g_1 n_{c1} + \bar g_2 n_{c2} \pm \sqrt{ (\bar g_1 n_{c1} - \bar g_2 n_{c2})^2 +4 g_{12}^2 n_{c1} n_{c2} } \bigg].$$
\begin{equation} \label {Chmp}
\mu_{+,-}=  \frac{\bar g_1 n_{c1}} {2} f_{+,-} (\Delta, \alpha),
\end{equation}
where $f_{+,-} (\Delta, \alpha)= 1 + \alpha \pm \sqrt{ (1-\alpha)^2 +4 \Delta ^{-1}\alpha }$ and
$\alpha= \bar g_2 n_{c2}/\bar g_1 n_{c1}$.\\
In the limit $k \rightarrow 0$,  we have $\varepsilon_{jk}= \hbar c_j k$  where $c_j= \sqrt{\bar g_j n_{cj} /m_j}$ is the sound velocity of a single condensate. 
The total dispersion is phonon-like in this limit
\begin{equation} \label{sound}
\varepsilon_{k (+,-)}= \hbar c_{+,-} k,
\end{equation} 
where the sound velocities $c_{+,-}$  are
 \begin{equation} \label{sound1}
c_{+,-} ^2=\frac{1}{2} \left[ c_1^2+c_2^2 \pm \sqrt{ \left( c_1^2-c_2^2\right) ^2 + 4 \Delta^{-1} c_1^2 c_2^2} \right].
\end{equation}
For $g_{12}^2> \bar g_1\, \bar g_2$, the spectrum (\ref{Bog}) becomes unstable and thus, the two condensates spatially separate.
We can see that the sound velocity $c_{+,-} \rightarrow 0$ as $T\rightarrow T_c$ since $n_{cj}=\tilde m_j=0$ near the transition, 
which means that the phonons in the TDHFB theory are the soft modes of the Bose-condensed mixture. 
%One should stress that the finite temperature behavior of a mixture of Bose like superfluid $^4{}$He \cite{CFetter, YNep, Khal} is different from that of a Bose-Bose mixture.
%the observed excitation of two-roton pairs \cite{Wong} .
%\bibitem {Wong} K. Wong, Phys. Left., 30A, 292 (1969); 32A, 195 (1970); K. W. Wong and W. Meyer, Nuovo Cim.,Bll, 155 (1972).

\subsection{Quantum and thermal fluctuations}

A straightforward calculation using Eq.(\ref{Inv1}) permits us to rewrite the normal and anomalous densities in terms of $\sqrt{I_k}$ \cite{Boudj,Boudj4}
\begin{equation}\label {nor}
\tilde n_j=\frac{1}{2}\int \frac{d \bf k} {(2\pi)^3} \left[\frac{E_k+ \mu_{+,-}} {\varepsilon_{k (+,-)}} \sqrt{I_{kj}}-1\right],
\end{equation}
and
\begin{equation}\label {anom}
\tilde m_j=-\frac{1}{2}\int \frac{d \bf k} {(2\pi)^3} \frac{ \mu_{+,-} } {\varepsilon_{k (+,-)}} \sqrt{I_{kj}}.
\end{equation}

At $T=0$, the total  depletion $\tilde n=\tilde n_1+\tilde n_2$ can be calculated via the integral (\ref {nor})
\begin{equation}\label {nor1}
\tilde n= \frac{1}{6 \sqrt{2} \pi^2} \left(\frac{1}{\xi_+^3}+\frac{1}{\xi_-^3} \right),
\end{equation}
where $\xi_{+,-}=\hbar/\sqrt{m \mu_{+,-}}$.\\
As remarked in integral (\ref{anom}), dimensional analysis suggests that we face the ultraviolet divergences in the expression of $\tilde m$ as anticipated above.
This problem can be cured by means of the dimensional regularization \cite{Boudj, Zin, Klein, Anders}  which follows from perturbation theory of scattering.
It gives asymptotically exact results at weak interactions  (for more details, see Appendix A of \cite{Boudj}).
This yields for the total anomalous density $\tilde m=\tilde m_1+\tilde m_2$:
\begin{equation}\label {anom1} 
\tilde m= \frac{1}{2\sqrt{2} \pi^2} \left(\frac{1}{\xi_+^3}+\frac{1}{\xi_-^3} \right).
\end{equation}
Importantly, the above expressions of the noncondensed and anomalous densities are proportional to $g_j^2$ and $g_{12}^2$.
For $g_{12}=0$, Eqs.(\ref {nor1}) and (\ref{anom1}) recover those obtained by the second-order Beliaev theory \cite{Beleav} and the perturbative
time-dependent mean-field scheme \cite{Giorg}.

%Notice that $\tilde m$ can be also perfectly calculated using the renormalized coupling constants:
%i.e.  we add $ g_j ^2 \int d {\bf k} /E_k$ and $g_{12} ^2 \int d {\bf k} /E_k$ to the r.h.s of Eq.(\ref{anom}).
%This replacement is in fact equivalent to the self-consistent ladder diagram approximation
%for the $T$-matrix given in (\ref{Ren}) which includes automatically the effects originating from the anomalous density \cite{Burnet, Boudj2015}.

\begin{figure}
\centerline{
\includegraphics[scale=0.8]{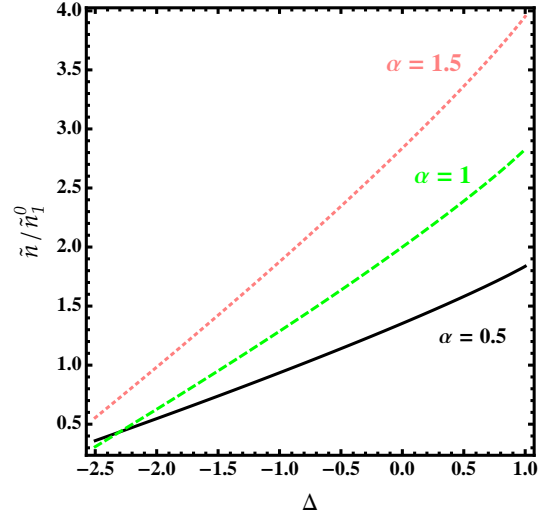}}
 \caption{(Color online) Condensed depletion  from Eq.(\ref{norT}) as a function of $\Delta$ for different values $\alpha$.}
\label{dd} 
\end{figure}

From now onward, we  assume that $\tilde m/ n_c \ll 1$,  this condition is valid at low
temperature and necessary for the diluteness of the system \cite{Yuk, Boudj2015}.
Therefore, the condensate depletion (\ref{nor1}) reduces to 
\begin{equation}\label {norT}
\tilde n= \frac{1}{2 \sqrt{2}} \tilde n_1^0 \left [f_+^{3/2} (\Delta, \alpha)+f_-^{3/2} (\Delta, \alpha) \right],
\end{equation}
%\tilde n= \tilde n_1+\tilde n_2= \frac{1}{3 \pi^2}  \left[ \frac{1} {\xi_1^3}+\frac{1} {\xi_2^3} \right],
where $\tilde n_1^0=(8/3) n_{c1}  \sqrt{n_{c1} a_1^3/\pi}$ is the single condensate depletion (type-1).
The depletion (\ref{norT}) is formally similar to that obtained by Tommasini et \textit {al}. \cite{Tom} 
using the Bogoliubov theory, with only $n_{cj}$ appearing as a corrected parameter instead of the total density $n_j$. 
At $T=0$ and for fixed density $\tilde n_1^0$,  the noncondensed fraction is proportional to $\Delta$ and $\alpha$
signaling that the number of excited atoms increases with $\Delta$ and $\alpha$ as is displayed in Fig.\ref{dd}.\\
The anomalous density of the mixture (\ref{anom1}) simplifies
\begin{equation}\label {anomT} 
\tilde m=\frac{1}{2 \sqrt{2}} \tilde m_1^0 \left [f_+^{3/2} (\Delta, \alpha)+f_-^{3/2} (\Delta, \alpha) \right],
\end{equation}
where $\tilde m_1^0= 8 n_{c1} \sqrt{n_{c1} a_1^3/\pi}$ is the anomalous density of a single component.
To the best of our knowledge, Eq.(\ref {anomT}) has never been derived in the literature.
It shows that $\tilde m$ is larger than $\tilde n$ similarly to the case of a single component. 
This indicates that the anomalous density is significant even at zero temperature in Bose-Bose mixtures.
We see also that $\tilde m$ is increasing with $(n_{c1} a^3)^{1/2}$, $\Delta$ and $\alpha$.
If the interspecies and intraspecies interactions were strong enough, the pair anomalous density becomes important results in
a large fraction of the total atoms would occupy the excited states.

At temperatures $T\ll g_1 n_{c1} $, the main contribution to integrals (\ref{nor}) and (\ref{anom}) comes from the long-wavelength region
where the spectrum takes the form (\ref{sound}).
Then the use of the integral $\int_0^ {\infty} {x^{2j-1}\left[\coth(\alpha x)-1\right]dx}=\pi^{2j}|B_{2j}|/ 2j\alpha^{2j}$ \cite{Yuk},
where $B_{2j}$ are the Bernoulli number, allows us to obtain the following expressions for the thermal
contribution of the noncondensed and anomalous densities:
\begin{align} \label {thfluc}
\tilde{n}_{th} &=|\tilde{m}_{th} | = \frac{2 \sqrt{2} }{3} n_{c1} \sqrt{\frac{ n_{c1} a^3}{\pi}} \left(\frac{\pi T} {n_{c1} g_1}\right)^2  \\
&\times \left [f_+^{-1/2} (\Delta, \alpha)+f_-^{-1/2} (\Delta, \alpha) \right]. \nonumber
\end{align}
Equation (\ref{thfluc}) shows clearly that $\tilde n$ and $\tilde m$ are of the same order of magnitude at low temperature and only their signs are opposite. 
A comparaison between Eqs.(\ref{norT}), (\ref{anomT}) and (\ref{thfluc}) shows that at $T\ll gn_c$, thermal fluctuations are smaller than
the quantum fluctuations.

Let us now discuss some relevant cases predicted by Eqs.(\ref{norT}) and (\ref{anomT}), 
for a balanced mixture where $n_1=n_2=n$, and $g_1=g_2=g_{12}$. Hence, the noncondensed and the anomalous densities reduce respectively to
$\tilde n= (4 /\sqrt{2}) \, \tilde n^0$ and $\tilde m=  (4 /\sqrt{2}) \, \tilde m^0$, whereas at low temperature, the thermal depletion and the anomalous density 
turn out to be given as $\tilde{n}_{th} =\tilde{m}_{th} =(2/3) n_{c1} \sqrt{n_{c1} a_1^3/\pi} \left(\pi T/n_{c1} g_1\right)^2$. 
Near the phase separation where $ g_{12}^2 \rightarrow \bar g_1 \bar g_2$, 
the condensate depletion and the anomalous density become, respectively $\tilde n= (1 + \alpha)^{3/2} \tilde n_1^0 $ and $\tilde m= (1 + \alpha)^{3/2} \tilde m_1^0 $.
At low temperature, the lower branch has the free-particle dispersion law: $\varepsilon_{k-} = E_k$ \cite {CFetter}  
while the upper branch is phonon-like $\varepsilon_{k+} =\hbar c_1 (1 + \alpha)^{1/2} k$. 
Therefore, the thermal depletion has a distinct temperature dependence as
\begin{align} \label {thfluc1}
\tilde n_{th} & = \frac{2 }{3} n_{c1} \sqrt{\frac{ n_{c1} a_1^3}{\pi}} (1 + \alpha)^{-1/2} \left(\frac{\pi T} {n_{c1} g_1}\right)^2  \\
&+ \left(\frac{m T} {2\pi \hbar^2}\right)^{3/2} \zeta (3/2), \nonumber 
\end{align}
where $\zeta (3/2)$ is the Riemann Zeta function. 
The second term in (\ref{thfluc1}) is the density of noncondensed atoms in a noninteracting gas.  
This reveals that the component associated with lower branch becomes ultradilute.
Notice that a similar temperature dependence distinction was obtained earlier by Colson and Fetter \cite {CFetter} for ${}^4$He-${}^6$He mixture. 
Such a distinction in the temperature dependence cannot occur in $\tilde m_{th}$ where the term $\propto T^{3/2}$ 
is absent since the anomalous density itself does not exist in an ideal gas \cite{Boudj, Griffin}.

%undergo instability for both cases.
%The instability manifests in the fact that $c_-$  becomes complex.

\subsection{Thermodynamics}

Corrections to the EoS of the mixture due to quantum and thermal fluctuations can be derived from Eq.(\ref{EoS}). 
Combining Eqs.(\ref{norT}), (\ref{anomT}) and (\ref{thfluc}) gives 
\begin{align} \label{EoS1}
\delta \mu &=\frac{\mu_1^0 }{2 \sqrt{2}}   \bigg\{ \left [f_+^{3/2} (\Delta, \alpha)+f_-^{3/2} (\Delta, \alpha) \right] \\
&+\frac{1}{2} \left [f_+^{-1/2} (\Delta, \alpha)+f_-^{-1/2} (\Delta, \alpha) \right] \left(\frac{\pi T} {n_{c1} g_1}\right)^2  \bigg\}, \nonumber
\end{align}
where $\mu_1^0= (32/3) g_1 n_{c1} \sqrt{n_{c1} a_1^3/\pi}$ is the zero temperature chemical potential of a single condensate.
At zero temperature and for $g_{12}=0$, Eq.(\ref{EoS1}) reduces to the seminal Lee-Huang-Yang (LHY) corrected EoS \cite{LHY} for 
one component BEC.

%Given the EoS, we can then compute the corrections to the inverse compressibility $ \delta \kappa^{-1} = n^2 \partial \delta \mu /\partial n$:
%\begin{equation}\label {compr}
%\delta \kappa^{-1}= \delta {\cal K}^{-1}  \left [f_+^{1/2} (\Delta, \alpha)+f_-^{1/2} (\Delta, \alpha) \right] \\
%\end{equation}
%where  $\delta {\cal K}^{-1}=16\,g_1 n_{c1}^2 \sqrt{n_{c1} a^3/\pi} $ is the single condensate compressibility.

At finite temperature, the grand-canonical ground state energy can be calculated using  the thermodynamic relation
$E=E_0+\delta E=-T^2 \left( \frac{\partial} {\partial T } \frac{F}{T} \right) |_{V,\mu}$ where the free energy is given by 
$F=E+T\sum_{\bf k}\ln[1-\exp(-\varepsilon_{k(+,-)}/T)]$, and 
$E_0=(g_j/2) \sum_j ( n_{cj}^2+ 4 n_{cj} \tilde n_j +2\tilde n_j^2 +|\tilde m_j|^2 + 2 n_{cj} \tilde m_j) + g_{12}  n_1 n_2$. 
When  $\tilde m_j/n_{cj} \ll1$ and $\tilde n_j/n_{cj} \ll1$, one has $E_0=(g_j/2) \sum_j n_{cj}^2 + g_{12}  n_{c1} n_{c2}$. 
The shift to the ground state energy due to quantum and thermal fluctuations is defined as 
\begin{align} \label{CGSE}
\delta E&= \frac{1}{2} \sum_j \bigg[\sum_k \left[\varepsilon_{k(+,-)} -( E_k+\mu_{+,-}) \right] \\
&+\sum_k \varepsilon_{k(+,-)}  \left(\sqrt{I_{kj}}-1 \right) \bigg]. \nonumber
\end{align}
The first term on the r.h.s of (\ref{CGSE}) which represents the energy corrections due to quantum fluctuations is 
ultraviolet divergent. To circumvent such a divergency, we will use the standard dimensional regularization.
The second term accounts for the thermal fluctuation contributions to the energy. The main contribution to it comes from the phonon region.
After some algebra, we obtain
\begin{align} \label{GSE}
\delta E&=\frac{1}{4\sqrt{2}}  E_1^0  \bigg\{ \left [f_+^{5/2} (\Delta, \alpha)+f_-^{5/2} (\Delta, \alpha) \right] \\
&+\frac{1}{2\sqrt{2}} \left [f_+^{-3/2} (\Delta, \alpha)+f_-^{-3/2} (\Delta, \alpha) \right] \left(\frac{\pi T} {n_{c1} g_1}\right)^4  \bigg\}, \nonumber
\end{align}
where $E_1^0/V= (64/15) g_1 n_{c1}^2 \sqrt{ n_{c1} a_1^3/\pi}$ is the zero-temperature single condensate ground state energy which can be obtained also
by integrating the chemical potential with respect to the density.
The same result could be obtained within the renormalization of coupling constants which consists of
adding $ g_j ^2 \int d {\bf k} /E_k$ and $g_{12} ^2 \int d {\bf k} /E_k$ \cite{Larsen, Petrov} to the r.h.s of Eq.(\ref{CGSE}).

In the case $g_{12}=0$, we read off from (\ref{GSE}) that $\delta E$ reduces to the ground-state energy of a single Bose gas. 
At $T=0$ and for $n_c=n$, $\delta E$ becomes identical to the Larsan's formula \cite{Larsen}.
Equation (\ref{GSE}) is a finite-temperature extension of that recently obtained by Cappellaro et \textit{al}. \cite{Capp} for a balanced mixture using 
the functional integration formalism within a regularization of divergent Gaussian fluctuations.
Indeed, the resulting ground-state energy is appealing since it furnishes an extra repulsive term proportional to $n_c^{5/2}+ n_c^{-3/2} T^4$
balancing the attractive mean-field term, allowing quantum and thermal fluctuations to stabilize  mixture droplets at finite temperature.
Quantum stabilization and the related droplet nucleation was proposed in Bose-Bose mixtures with \cite{Petrov, PetAst, Cab, Sem} and without Rabi coupling \cite{Capp} 
as well as applied on dipolar condensates \cite {Pfau,Wach, Bess, Chom}.
The finite-temperature generalization of these LHY corrections to the case of a dipolar Bose gas has been also analyzed in our recent work \cite{BoudjDp}.

\section{Conclusion and outlook} \label{concl}

In this paper we have systematically studied effects of quantum and thermal fluctuations on the dynamics and the collective excitations of 
a two-component Bose gas utilizing the TDHFB theory.
%Within the time-dependent BV variational principle we have derived a set of coupled equations describing
%the spatiotemporal evolution of the two condensates, the thermal clouds and the anomalous averages together in an integrated way.
We revealed that our approach is able to capture the qualitative evolution of two-component BECs at finite temperature.
The impact of the anomalous fluctuations on the miscibility criterion for the mixture was discussed.

Within an appropriate numerical method,  we elucidated the behavior of the condensed, the noncondensed and the anomalous densities in terms of temperature 
for both miscible and immiscible mixtures under spherical harmonic confinement.
We demonstrated in particular that the mixture undergoes a transition from miscible to immiscible regime 
owing to the predominant contribution of anomalous fluctuations notably at $T\simeq 0.5 T_c$.
We found  that such fluctuations are also the agent responsible for the strong damping of the relative motion of the centers-of-mass of the condensed and thermal components.
At zero temperature, the TDHFB results correct the existing theoretical models, making the theory in good agreement with JILA experiment \cite{Hall}.

We linearized our TDHFB equations using  the RPA for a weakly interacting uniform Bose-Bose mixture.
This method is a finite-temperature extension of the famous second-order Beliaev approximation \cite {Beleav}.
The TDHFB-RPA theory provides us with analytical machinery powerful enough to calculate quantum and thermal 
fluctuation corrections to the excitations, the sound velocity, the EoS and the ground-state energy. 
We compared the theory with previous theoretical treatments and excellent agreement has been found in the limit $\tilde m/n_c \ll 1$.
One should stress that the results of our TDHFB-RPA technique can be generalized to the case of a harmonically trapped mixture using the local density approximation. 

The findings of this work are appealing for investigating the properties of mixture droplets at finite temperature.
An important topic for future work is to look at how thermal fluctuations manifest themselves in dipolar Bose-Bose mixtures.

\section{Acknowledgements}
We are indebted to Eugene Zaremba and Hui Hu for stimulating discussions.

\end{document}